# When Local Governments' Stay-at-Home Orders Meet the White House's "Opening Up America Again"

**Reza Mousavi**
University of Virginia
mousavi@virginia.edu

**Bin Gu**
Boston University
bgu@bu.edu

## Abstract

On April 16th, The White House launched "Opening up America Again" (OuAA) campaign while many U.S. counties had stay-at-home orders in place. We created a panel data set of 1,563 U.S. counties to study the impact of U.S. counties' stay-at-home orders on community mobility before and after The White House's campaign to reopen the country. Our results suggest that before the OuAA campaign stay-at-home orders brought down time spent in retail and recreation businesses by about 27% for typical conservative and liberal counties. However, after the launch of OuAA campaign, the time spent at retail and recreational businesses in a typical conservative county increased significantly more than in liberal counties (15% increase in a typical conservative county Vs. 9% increase in a typical liberal county). We also found that in conservative counties with stay-at-home orders in place, time spent at retail and recreational businesses increased less than that of conservative counties without stay-at-home orders. These findings illuminate to what extent residents' political ideology could determine to what extent they follow local orders and to what extent the White Houses' OuAA campaign polarized the obedience between liberal and conservative counties. The silver lining in our study is that even when the federal government was reopening the country, the local authorities that enforced stay-at-home restrictions were to some extent effective.

## Introduction:

The COVID-19 pandemic has been raging in the world since December 2019. The United States reported its first COVID-19 patient in January 21st, 2020.[1] As of August 12, 2020 there are over 5 million confirmed cases in U.S. with a death toll of over 165,000.[2] The immense impact of COVID-19 pandemic on the lives of billions of people has forced authorities to devise response strategies to contain the damage. From enforcing stay-at-home orders to restrict public and private gatherings to wearing masks and social distancing, the authorities have come up with a variety of response strategies. These restrictive measures are effective only if they are adhered to by citizens. Given the diverse set of opinions held by citizens, their level of adherence to the restrictions could be different. Results from initial research studies provide some evidence of these differences based on residents' political beliefs. For instance, Painter and Qiu (2020) found that residents' political beliefs affects their compliance with social distancing orders that are imposed as a response to the spread of COVID-19. Their findings reveal that counties that voted for President Trump in 2016 presidential election are more likely to disobey the social distancing orders. Another study by Grossman et al. (2020) also revealed that the Democratic counties were more likely to obey stay-at-home orders enforced by the state governors. They also found that Democratic counties with Republican governors are more likely to stay at home when compared to other counties. Overall, the current research signals the impact of stay-at-home orders on residents' stay-at-home behaviors. Although both studies offer similar findings, they both use data from SafeGraph's shelter-in-place data set.[3] This data set uses a sample of users' cell phone locations to determine the time stayed at home. However, we believe that another data set published by Google would be more informative as it not only includes data about stay-at-home time, but also it includes data about the amount of time users spend in retail & recreation locations (such as bars, restaurants, and gyms), grocery stores, parks, places of work, and transit locations. We believe that such data would be more informative in measuring the impact of stay-at-home orders.

---

[1] https://abcnews.go.com/Health/timeline-coronavirus-started/story?id=69435165
[2] https://coronavirus.jhu.edu/map.html
[3] https://www.safegraph.com/dashboard/covid19-shelter-in-place?s=US&d=08-20-2020&t=counties&m=index

Another distinction between our research and prior studies is about the federal government's intervention: "[w]e are starting our life *AGAIN!*,"[4] said the U.S. president during his Coronavirus Task Force press briefing on April 16th. A day earlier, during another press briefing, the president claimed that the U.S. has "passed the pick on new cases."[5] Yet a couple of days before that, on April 13th, he also claimed "total authority"[6] over governors. The three Coronavirus Task Force press briefings on April 13th, 15th, and 16th mark a major shift in U.S. president's response policy regarding the pandemic. A new policy that was also cascaded in president's tweets (see Figure 1). What followed this new policy was a set of guidelines[7] to open up America. The White House launched "Opening up America Again" (OuAA) website on Arpil 16th. Along with these guidelines, president Trump used his powerful Twitter account to encourage protestors (mainly composed of his supporters) to "liberate" Michigan and Minnesota, two states with Democratic governors who imposed strict social distancing restrictions (Shear and Mervosh 2020).

---

[4] https://www.whitehouse.gov/briefings-statements/remarks-president-trump-vice-president-pence-members-coronavirus-task-force-press-briefing-27/

[5] https://www.whitehouse.gov/briefings-statements/remarks-president-trump-vice-president-pence-members-coronavirus-task-force-press-briefing-26/

[6] https://www.whitehouse.gov/briefings-statements/remarks-president-trump-vice-president-pence-members-coronavirus-task-force-press-briefing-25/

[7] https://www.whitehouse.gov/openingamerica/

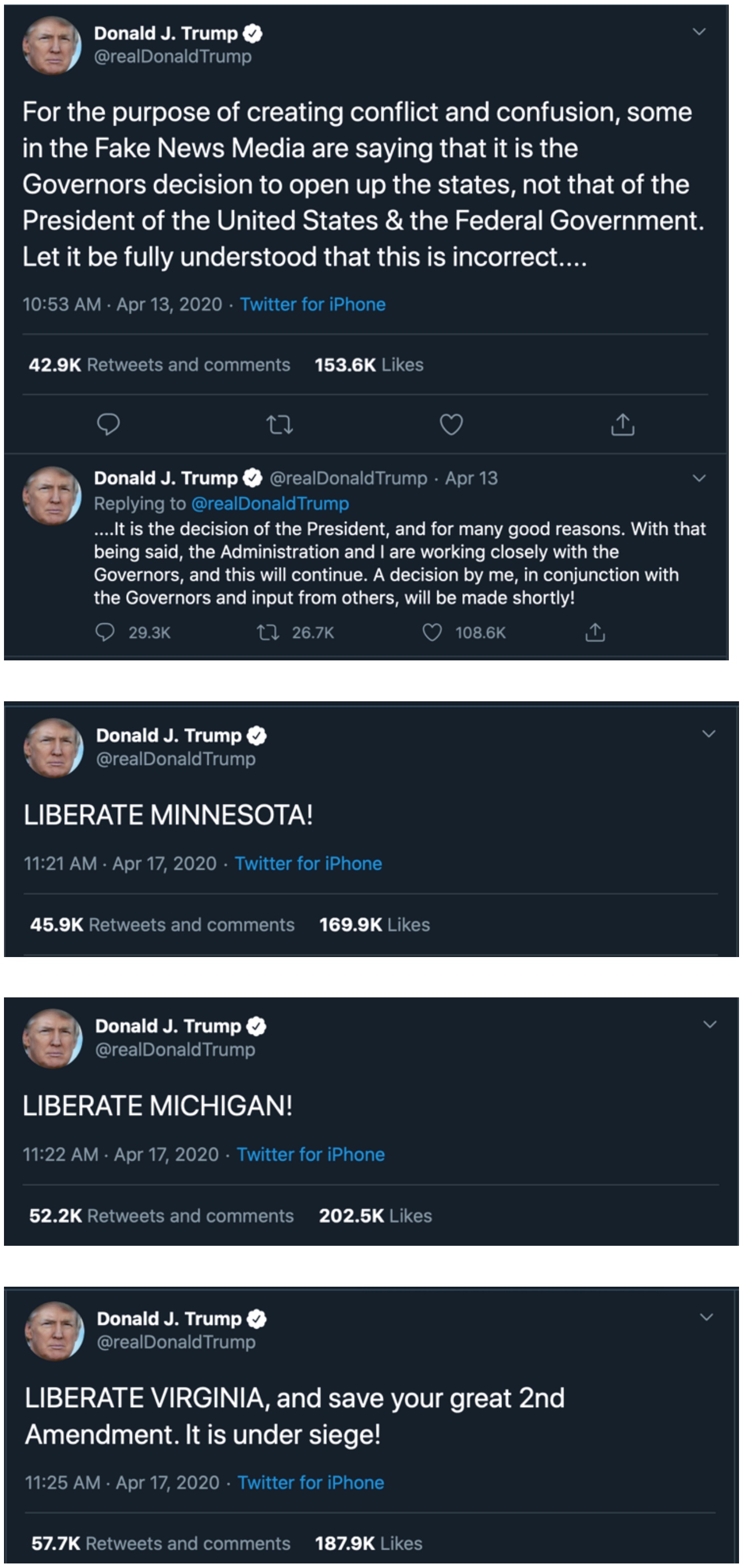

**Figure 1. President Trump's Tweets about his "Opening Up America Again" Campaign**

In this study, we use weekly panel data about U.S. counties' community mobility, unemployment rate, political orientation and COVID-19 cases and deaths along with stay-at-home and shelter-in-place restrictions to understand the impact of stay-at-home orders on community mobility and to what extent this impact is moderated by the political orientation of the county and by the OuAA campaign. Our findings reveal that:

1- Stay-at-home and shelter-in-place restrictions imposed by counties and states decreased time spent at retail & recreation places such as bars, restaurants, gyms, and movie theatres, and increased time spent at residential places.
2- Liberal counties spent more time at home and less time at retail stores compared to conservative counties during the stay-at-home orders.
3- Liberal counties spent more time at home and less time at retail stores compared to conservative counties after OuAA campaign launched by the White House.
4- Conservative counties that had a stay-at-home order in place spent more time at home and less at retail & recreation places even after OuAA campaign launch when compared to conservative counties that did not have stay-at-home order in place.

Our results are based on a quasi-experimental setting and we have controlled for the number of cases and deaths per 100k population, county fixed effects, time fixed effects and the interaction of state and time fixed effects. We also examined the robustness of our findings by running the models using a matched sample of counties. We used county level data about residents' education, population, deaths, births, domestic and international migration, percent below federal poverty line, unemployment rate, and median household income. In this summary, we report our study design, data analysis, and preliminary findings.

## Data:

We created a panel data set of U.S. counties by collecting data about community mobility scores, COVID-19 new cases and deaths (adjusted by population), ideological orientation of counties, and state's COVID-19 response data (stay-at-home, shelter-in-place, and other types of restrictions and their timelines) for the period between the first week of March 2020 and the first week of June 2020.

From 3,141 U.S. counties, we removed any county that did not have a value for the variables in our models. For instance, many counties did not have a value for the community mobility indices (Retail and Residential described in table 1). After removing the counties with missing values, we ended up with panel data of 1,563 counties observed during a 14-week period. Below, we describe the sources of data and our data collection approach:

1- *Community Mobility Data:*

This data set was obtained from Google.[8] Google's community mobility dataset "shows how visits to places, such as grocery stores and parks, are changing in each geographic region."[9] According to the documentations, this dataset shows how visits and length of stay at different places change compared to a baseline. The baseline is the median value, for the corresponding day of the week, during the 5-week period from January 3rd, 2020 to February 6th, 2020. Google indicates that the data is included in the calculations based on user settings, connectivity, and whether there is any privacy concern (due to small sample size in some areas). If there are any concerns regarding the privacy of Google's users, the data fields will be left empty. Due to these omitted values, we did not include counties with missing values.

Google's community mobility data set includes six categories of places: grocery & pharmacy, parks, transit stations, retail & recreation, residential, and workplaces. From these six categories, we used retail & recreation and residential categories. The reason why we limited our study to these two categories is that these two categories portray a more accurate picture of community mobility trends during the pandemic. Retail & recreation category includes mobility trends for places like restaurants, cafes, shopping centers, theme parks, museums, libraries, and movie theatres. These businesses are non-essential businesses that could be avoided during the pandemic. If stay-at-home orders are effective, we would expect a decrease in retail & recreation trend. Residential category refers to the mobility trends of places of residence. If stay-at-home orders are effective, we would expect an upward trend in residential category. Other categories such as grocery & pharmacy stores, workplaces, and transit stations are either essential or determined by the employers rather than the residents themselves. Therefore, trends in grocery & pharmacy stores, workplaces, and transit stations may not provide us with a reasonable pattern about residents' will in adhering to stay-at-home orders and other types of restrictions. We also excluded parks mobility trends because visiting such places would be possible with very limited risk of infection (people could stay six feet apart in the open areas). With that said, we created two dependent variables *Retail* and *Residential* based on Google's mobility trends for retail & recreations and residential places respectively.

---

[8] https://www.google.com/covid19/mobility/

[9] https://www.google.com/covid19/mobility/data_documentation.html?hl=en

2- *COVID-19 Data:*

This data set was obtained from NY Times' GitHub page.[10] This dataset includes the number of new cases and the number of new deaths per day per county per 100k residents. We aggregated this data by taking the means of daily new cases and daily new deaths over each week.

3- *Ideological Orientation Data:*

This data set was obtained from American Ideology Project.[11] We used the 2016 release of "County-Level Preference Estimates". From this data file, we used variable mrp_mean, which is the estimate of the mean ideology of the county. This measure ranges from a negative number to a positive number. The smallest value in mrp_mean represents the most liberal county, while the largest value in mrp_mean is associated with the most conservative county. Therefore, we can interpret mrp-_mean as a metric for gauging the level of conservativeness of a county. Hence, in our study, we call this variable *conservative*. In the 2016 release of the data, *conservative* ranges from -1.098 (the least conservative county) to 0.842 (the most conservative county). We used Min-Max transformation to transform the scale to range from zero (most liberal) to 1 (most conservative). The methodology for estimating ideological orientation scores is described in Tausanovitch and Warshaw (2013).

[10] https://github.com/nytimes/covid-19-data
[11] https://americanideologyproject.com

4- *County-level Restrictions Data:*

This data set was obtained from a GitHub repository.[12] The data summaries and the methods used for assembling the data sets are detailed in Killeen et al. (2020). There are multiple data files in the data repository, from which we used "interventions.csv" data file. This data file contains the dates that counties (or states governing them) enforced policies (such as stay-at-home orders) to mitigate the spread of COVID-19 by restricting community mobility or gatherings. In addition to the dates of policy enforcement initiations, this data set includes the dates the polices were rolled back. From this data set, we used five types of restrictions:

1- Type 1: Stay-at-home orders

2- Type 2: Prohibiting gatherings of 50 or more people

3- Type 3: Prohibiting gatherings of 500 or more people

4- Type 4: Prohibiting dine-in restaurants and bars

5- Type 5: Closing entertainment businesses and gyms

Type 1 is the main variable of interest in our study. We use Type 2 through Type 5 as control variables in our models.

6- *County-level socio-economic data:*

This data set was obtained from The U.S. Department of Agriculture (USDA).[13] and includes information about the socio-economic indicators at the county level. In particular, this data set includes information about education level, population estimates including national and international migration, poverty, and unemployment.

---

[12] https://github.com/JieYingWu/COVID-19_US_County-level_Summaries

[13] https://www.ers.usda.gov/data-products/county-level-data-sets/download-data/

*Variables:*

Table 1 reports the list of variables, their descriptions, and their summary statistics. Figure 2 reports the correlation coefficients. Our data set includes data about 1,563 counties over a 14-week period (from the 10th week of 2020 through the 23rd week of 2020). This is a period that covers the first peak in the number of cases in the U.S. and includes the time before and after the roll out of the OuAA campaign by The White House. Furthermore, many counties and states enforced stay-at-home orders during this time period. Some of those restrictions were lifted again in the same time frame of our study.

We excluded observations with any missing value from the data set. This resulted in 18,769 observations with complete data. As mentioned before, the community mobility indices (*Retail* and *Residential*) were obtained from Google's Community Mobility Report. According to Google, these data are based on "data from users who have opted-in to Location History for their Google Account".[14] *Retail* and *Residential* indices reflect the change in users' locations based on a baseline. Per Google's documentation, the baseline is the median value, for the corresponding day of the week, during the 5-week period from January 3rd through February 6th of 2020. A negative value for *Retail* means that users spent less time in retail stores compared to the baseline timeframe. A positive score for *Residential* means that the users spent more time at a residential location (i.e. home) compared to the baseline timeframe.

**Table 1. Descriptive Statistics of the Variables**

| Variable | Description | Count | Mean | Std. | Min | Max |
|---|---|---|---|---|---|---|
| Dependent Variables:[15] | | | | | | |
| Retail | Mobility trends for places like restaurants, cafes, shopping centers, theme parks, museums, libraries, and movie theaters. | 18,769 | -18.420 | 22.970 | -63.023 | 24.143 |

[14] https://www.google.com/covid19/mobility/data_documentation.html?hl=en

[15] Dependent variables are floored (at 0.005) and capped (at 0.995) to treat the outliers.

| | | | | | | |
|---|---|---|---|---|---|---|
| Residential | Mobility trends for places of residence. | 18,769 | 11.050 | 7.395 | -2.220 | 27.000 |
| *Independent Variables:* | | | | | | |
| stay | Whether the county enforced a stay-at-home order during the focal week. | Not Enforced: 9,288<br>Enforced: 9,481 | | | | |
| post_reopen | Equals 1 on or after week of April 13th and zero otherwise. | Before OuAA: 8,365<br>After OuAA: 10,404 | | | | |
| conservative | Ideological orientation of each county. This variable measures to what extent a county is conservative | 18,769 | 0.657 | 0.148 | 0 | 1 |
| *Control Variables:* | | | | | | |
| cases[16] | Number of daily new cases per 100k residents averaged per week | 18,769 | 3.941 | 5.741 | -14.670 | 389.368 |
| deaths | Number of daily new deaths per 100k residents averaged per week | 18,769 | 0.178 | 0.305 | -1.030 | 8.993 |
| unemployment | Weekly unemployment rate in the county | 18,769 | 9.825 | 5.529 | 1.800 | 34.300 |
| gathering50 | Whether gatherings of 50 or more were banned in the county during the focal week. | Not Enforced: 4,518<br>Enforced: 14,251 | | | | |
| gathering500 | Whether gatherings of 500 or more were banned in the county during the focal week. | Not Enforced: 3,664<br>Enforced: 15,105 | | | | |
| dine_in | Whether restaurant dine-ins were prohibited in the county during the focal week. | Not Enforced: 6,722<br>Enforced: 12,047 | | | | |
| gym | Whether entertainment businesses and gyms were closed in the county during the focal week. | Not Enforced: 6,689<br>Enforced: 12,080 | | | | |

[16] In some rare cases, the number of daily cases and deaths are negative. This is because of the adjustments made to the counts made for the previous days.

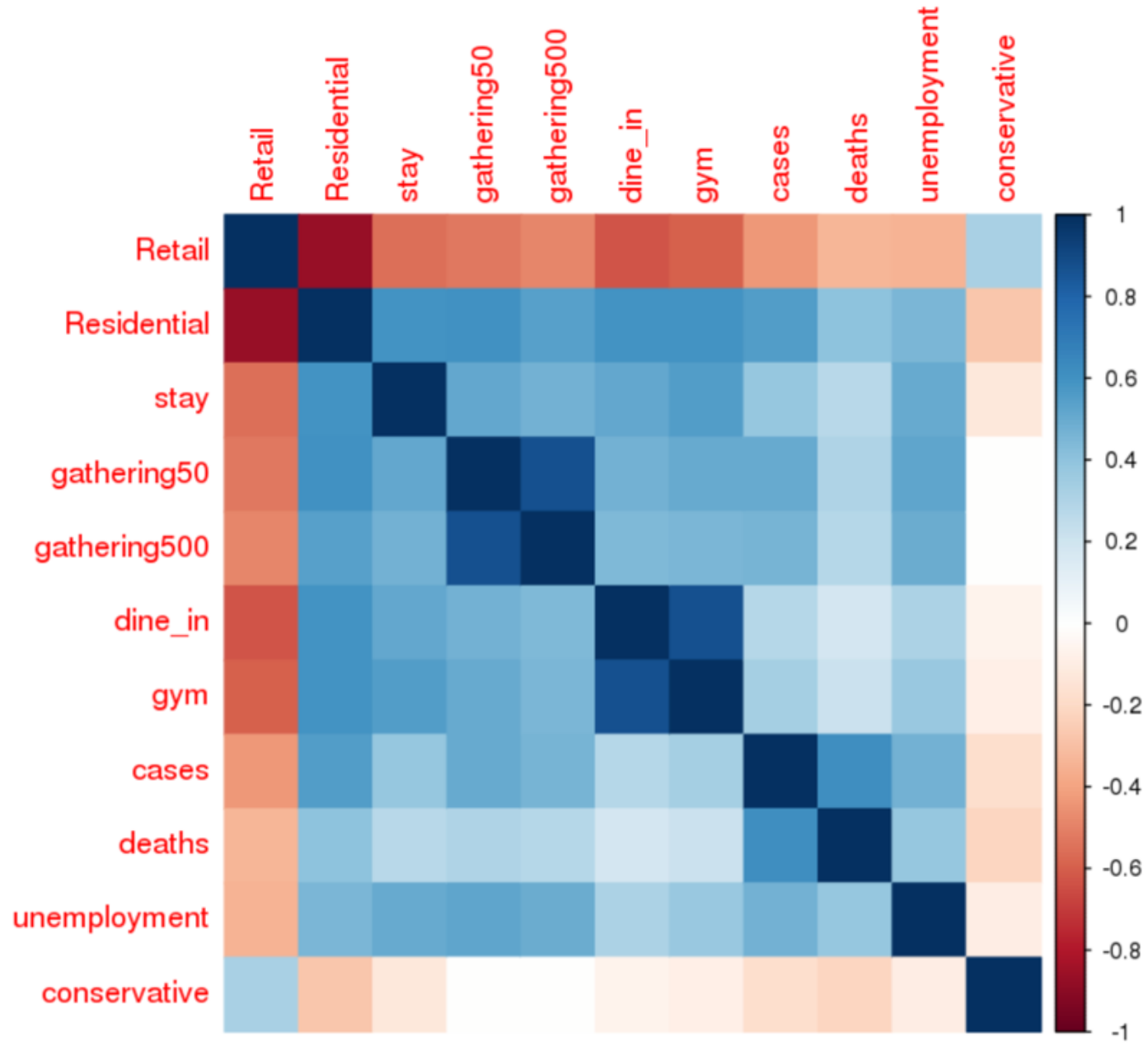

**Figure 2. Correlation Matrix (Correlation Coefficients)**

Figure 3 visualizes the longevity of stay-at-home orders during our study period. Lighter colors mean that the stay-at-home orders were in place for a short period of time (or not enforced at all), while darker colors represent longer stay-at-home orders. This plot only includes the counties we used in our analysis.

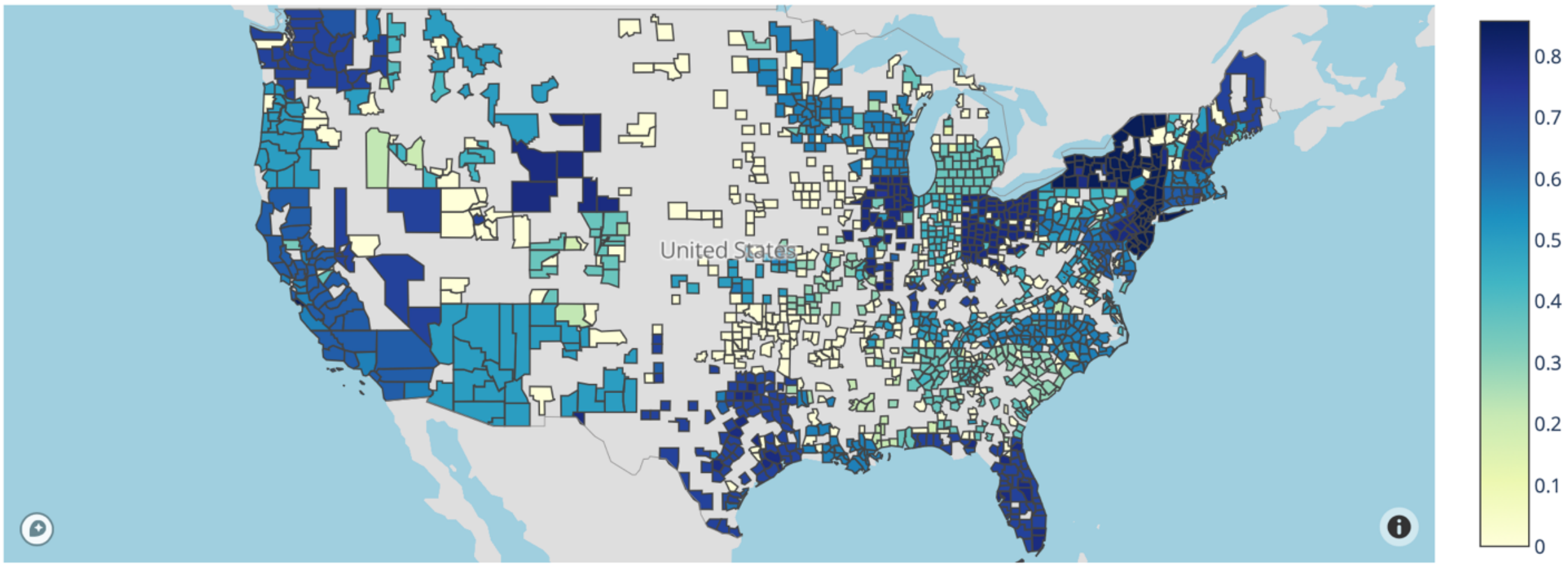

**Figure 3. Average County-level Stay-at-home Restrictions Color-coded Based on Longevity of Orders (counties with darker colors enforced stay-at-home orders for a longer period)**

Figure 4 compares the length of stay-at-home restrictions in U.S. states. The bars represent the week at least one county within the state enforced a stay-at-home order. New York, California, and New Jersey are among the states with longer duration of stay-at-home orders at least in one of their counties.

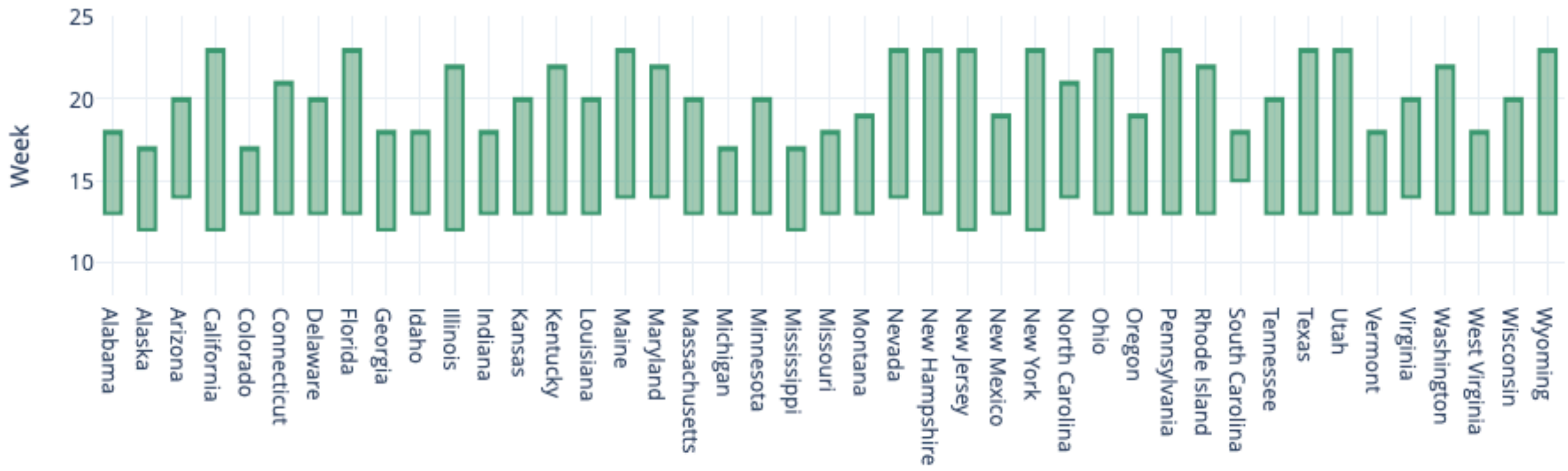

**Figure 4. Duration of stay-at-home Restrictions in U.S. States**

The two plots in Figure 5 visualize the average weekly trend in *Retail* (a) and *Residential* (b) indices over the period of our study. We separated the counties based on their ideological orientation (i.e. *conservative*). The counties with above median score for *conservative* are labeled as Conservative Counties and counties with score below the median for *conservative* are labeled as Liberal Counties. According to the plots Conservative Counties spent more time at retail locations and less time at residential locations. Also, we can observe that the retail activity in both Conservative Counties and Liberal Counties started to increase on week 16. We can also observe that both Conservative Counties and Liberal Counties spent less time at residential locations after OuAA campaign.

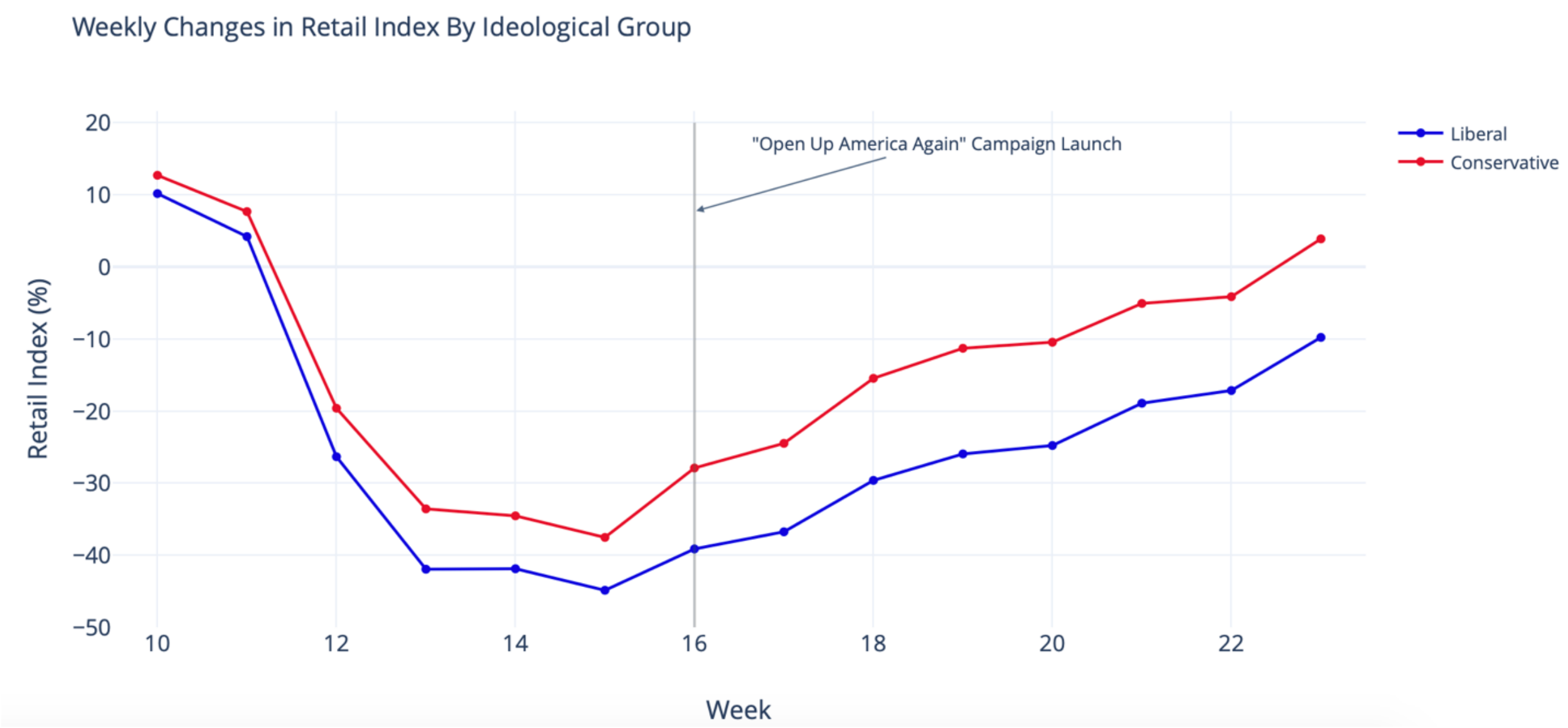

**(a) Retail Index**

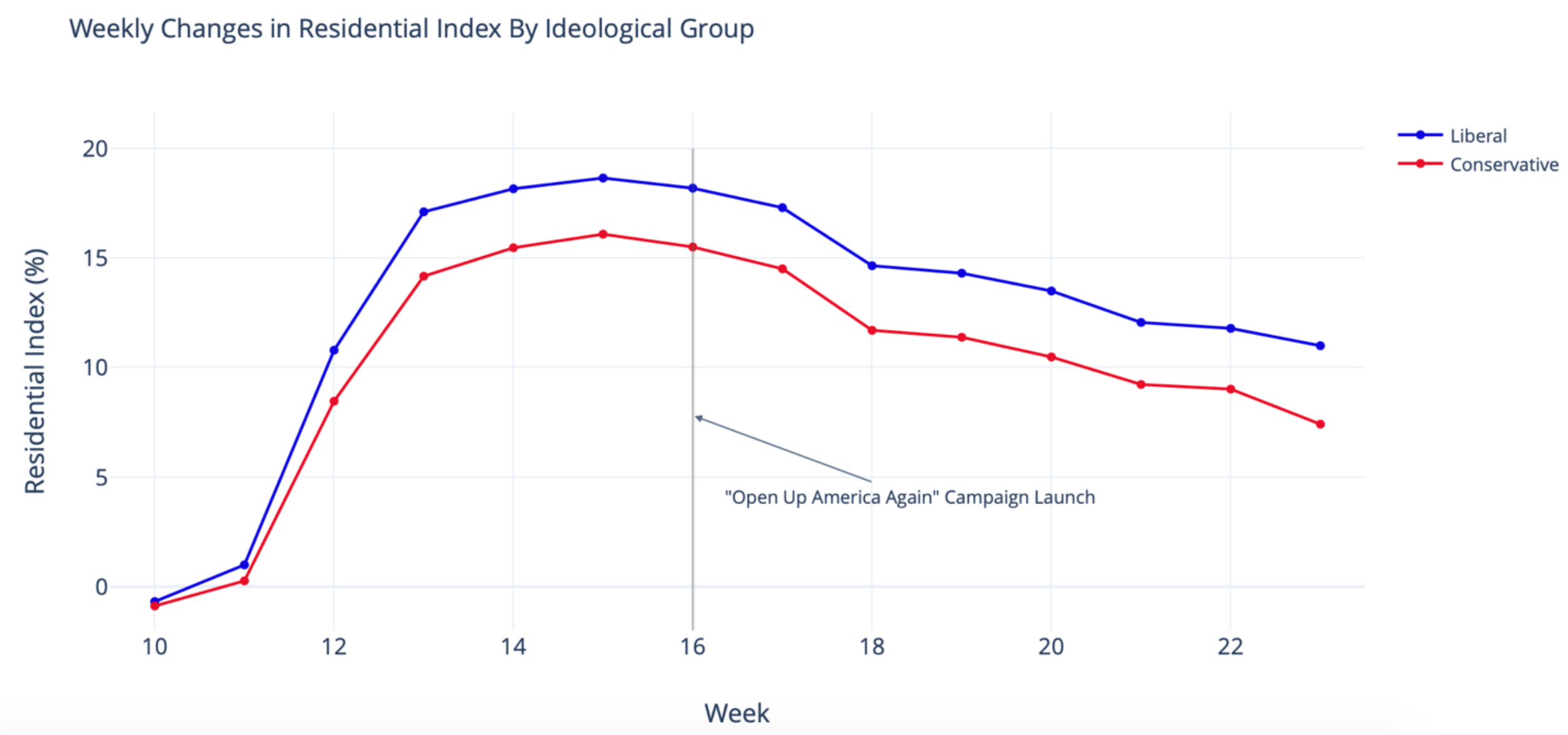

**(b) Residential Index**

**Figure 5. Weekly Changes in Retail and Residential Indices by Ideological Groups**

Before we introduce our econometric model, we present model-free comparisons of *Retail* and *Residential* indices in U.S. counties based on stay-at-home orders, counties' ideological category (Conservative County or Liberal County) and OuAA campaign. According to Table 2, in Liberal counties (Conservative County = 0) that did not have stay-at-home order in place (stay-at-home

order = 0), the value of *Retail* changed from -29.250 to -15.637 (increase in retail & recreation activity) and the value for *Residential* decreased from 11.988 to 11.750 after the launch of OuAA campaign. For Conservative counties without stay-at-home order the value for *Retail* changed from -24.656 to -4.126 and the value for *Residential* changed from 10.763 to 9.349. In Liberal counties that with stay-at-home order in place, the value of *Retail* changed from -43.025 to -29.852 (increase in retail & recreation activity) and the value for *Residential* decreased from 18.028 to 15.183 after the launch of OuAA campaign. For Conservative counties without stay-at-home order, the value for *Retail* changed from -37.070 to -17.063 and the value for *Residential* changed from 15.898 to 12.318 after the launch of OuAA campaign.

**Table 2. Model-free Comparison of U.S. Counties**

| Stay-at-home Order | Conservative County | After OuAA Campaign | Retail | Residential |
|---|---|---|---|---|
| 0 | 0 | 0 | -29.250 | 11.988 |
| 0 | 0 | 1 | -15.637 | 11.750 |
| 0 | 1 | 0 | -24.656 | 10.763 |
| 0 | 1 | 1 | -4.126 | 9.349 |
| 1 | 0 | 0 | -43.025 | 18.028 |
| 1 | 0 | 1 | -29.852 | 15.183 |
| 1 | 1 | 0 | -37.070 | 15.898 |
| 1 | 1 | 1 | -17.063 | 12.318 |

## Econometric Model:

We use the Difference-in-Difference (DiD) study design to understand the impact of stay-at-home orders on community mobility. The enforcement of stay-at-home orders by counties over time creates a natural experiment setting that allows the comparison of difference in community mobility before and after enforcing the stay-at-home orders across the counties. We further address the endogeneity of stay-at-home order decisions using a matched sample of counties (a match between counties that enforced an order and counties that never did). To assess the effect of stay-at-home orders on community mobility indices, we employ the following model:

$$y_{ist} = \alpha + \beta_0 \; stay_{ist}$$

$$+\beta_1 \; stay_{ist} \times conservative_{is} + \beta_2 \; post_reopen_t \times conservative_{is} + \beta_3 \; stay_{ist} \times post_reopen_t$$

$$+\beta_4 \; stay_{ist} \times post_reopen_t \times conservative_{is}$$

$$+\beta_5 \; cases_{ist} + \beta_6 \; deaths_{ist} + \beta_7 \; unemployment_{ist} + \beta_8 \; gathering50_{ist} + \; \beta_9 \; gathering500_{ist}$$

$$+\beta_{10} \; dine_in_{ist} + \beta_{11} \; gym_{ist}$$

$$+ \; \delta_t + \zeta_{is} + \delta_t \times \xi_s + \epsilon_{ist},$$

where *i* represents the county, *s* represents the state, and *t* represents the week. $y_{ist}$ is the community mobility index (i.e. *Retail* or *Residential*). We are interested in $\beta_0$ through $\beta_4$. $\beta_0$ is the DiD coefficient and $\beta_1$ through $\beta_4$ capture the interaction effects between *stay* and *conservative*, *post_reopen* and *conservative*, and *stay*, *post_reopen*, and *conservative* respectively. $\beta_5$ through $\beta_{11}$ capture the effects of the control variables. $\delta_t$ captures time-fixed effects, $\zeta_{is}$ captures county-fixed effects, and $\delta_t \times \xi_s$ capture the interaction between time- and state-fixed effects.

## Results:

Table 3 reports the results of our DiD analysis. For stay-at-home orders to be effective, we expect a drop in *Retail* and a jump in *Residential*. In models 1 through 3, the coefficient for *stay* is negative and significant. That is, retail & recreation activities decreased in counties that had a stay-at-home order after controlling for number of cases and deaths per 100k, unemployment rate, other types of restrictions, county-fixed effects, time-fixed effects, and the interaction of time- and state-fixed effects. In models 4 through 6, the coefficient for *stay* is positive and significant. This indicates that stay-at-home orders were effective in keeping residents at home. In model 2, the interaction between *stay* and *conservative* is positive and significant. This means that more conservative counties had more retail & recreation activities than liberal counties. In model 5 the coefficient for this interaction is negative and significant. This indicates that the conservative counties spent less time at home during the stay-at-home enforcement. In model 3, the interaction between *post_reopen* and *conservative* is positive and significant, which indicates that the conservative counties had more retail & recreation activities than liberal counties after OuAA campaign launch. This coefficient is negative and significant in model 6, suggesting more conservative counties stayed less at home after OuAA campaign launch compared to less conservative counties. In model 3, the interaction between *stay, post_reopen*, and *conservative* is negative and significant. This means that conservative counties that had a stay-at-home order enforced had less retail & recreation activities after OuAA launch compared to conservative counties that did not. The three-way interaction coefficient in model 6 also suggest that the conservative counties with stay-at-home order in place spent more time at home after the launch of OuAA campaign compared to conservative counties that did not.

**Table 3. The Impact of "stay-at-home" orders and "Reopen America" on Community Mobility Indices in U.S. Counties**

| | Retail | | | Residential | | |
|---|---|---|---|---|---|---|
| | Model 1 | Model 2 | Model 3 | Model 4 | Model 5 | Model 6 |
| stay | -2.590***<br>[0.475] | -3.172***<br>[0.475] | -2.687***<br>[0.625] | 1.137***<br>[0.123] | 3.118***<br>[0.162] | 4.595***<br>[0.223] |
| stay × conservative | | 7.630***<br>[0.642] | 6.866***<br>[0.887] | | -3.085***<br>[0.165] | -4.916***<br>[0.234] |
| post_reopen × conservative | | | 26.573***<br>[0.852] | | | -5.482***<br>[0.224] |
| stay × post_reopen | | | 4.677***<br>[1.157] | | | -3.723 ***<br>[0.305] |
| stay × post_reopen × conservative | | | -10.032***<br>[1.196] | | | 5.056***<br>[0.315] |
| cases | -0.051***<br>[0.005] | -0.049***<br>[0.005] | -0.052***<br>[0.005] | 0.019***<br>[0.001] | 0.018***<br>[0.001] | 0.018***<br>[0.001] |
| deaths | -0.535***<br>[0.111] | -0.449***<br>[0.111] | -0.307**<br>[0.106] | 0.229***<br>[0.029] | 0.194***<br>[0.028] | 0.180***<br>[0.028] |
| unemployment | -0.691***<br>[0.030] | -0.683***<br>[0.030] | -0.518***<br>[0.029] | 0.145***<br>[0.008] | 0.142***<br>[0.008] | 0.116***<br>[0.008] |
| other county restrictions | ✓ | ✓ | ✓ | ✓ | ✓ | ✓ |
| time fixed effects | ✓ | ✓ | ✓ | ✓ | ✓ | ✓ |
| county fixed effects | ✓ | ✓ | ✓ | ✓ | ✓ | ✓ |
| time fixed effects × state fixed effects | ✓ | ✓ | ✓ | ✓ | ✓ | ✓ |
| Observations | 18,769 | 18,769 | 18,769 | 18,769 | 18,769 | 18,769 |
| $R^2$ | 0.929 | 0.930 | 0.936 | 0.957 | 0.958 | 0.959 |
| F-statistic | 338.108*** | 340.659*** | 370.266*** | 573.525*** | 585.174*** | 604.519*** |

To better interpret the results, we can consider a typical conservative county[17] and a typical liberal county[18] that enforced stay-at-home order before OuAA's launch. For these two counties, we can define four phases over time:

- Phase 1: No stay-at-home order and before OuAA
- Phase 2: Stay-at-home order enforced and before OuAA
- Phase 3: Stay-at-home order enforced and after OuAA
- Phase 4: Stay-at-home order expired after OuAA.

For each one of these four phases, we used our model to obtain the predicted value for Retail and Residential for each one of those two counties. Figure 6 shows how these values change over each phase for a typical conservative county (a) and a typical liberal county (b). For instance, for this conservative county, retail & recreational activities decreased from almost zero to -27.360 once the county entered phase 2 (stay-at-home order enforced before OuAA). Once OuAA was launched, the county moved to phase 3 and retail & recreational activities increased to -12.290. When the county entered phase 4 (stay-at-home order lifted after launch of OuAA), retail activities increased to 3.484. For the liberal county in Figure 6 (b), the predicted value for Retail was -12.621. That is, even before the county enforced the stay-at-home order, the retail and recreational activities was 12.621 below the baseline (the 5-week period from January 3rd, 2020 to February 6th, 2020). When the county enforced the stay-at-home order (moved to phase 2), the predicted value for Retail decreased further to -40.553. That is, the retail and recreational activities dropped 27.931 points.

[17] We define a county with a score of one standard deviation (0.148) above the median (0.673) for variable *conservative* as a typical conservative county.
[18] We define a county with a score of one standard deviation (0.148) below the median (0.673) for variable *conservative* as a typical liberal county.

When OuAA was launched, the predicted value for Retail increased to -31.364 (9.189 points increase). Finally, when the county entered phase 4 (stay-at-home order expired), the predicted value of Retail further increased to -21.150 (10.214 points increase).

Overall, Figure 6 confirms that stay-at-home orders were to some extent effective in decreasing retail and recreational activities in both conservative and liberal counties. However, OuAA was more impactful in opening conservative counties than it was in liberal counties. In the example displayed in Figure 6, the conservative county's Retail score increased from -27.360 to -12.290 (15.070 points increase). However, the liberal county's Retail score increased from -40.553 to -31.364 (9.189 points increase). The difference in the increase in Retail is larger in the conservative county than in the liberal county (15.070 point Vs. 9.189). A similar pattern about the impact of OuAA can be observed when we consider Residential score. That is, the Residential score decreased more for the conservative county than it did for the liberal county after the launch of OuAA.

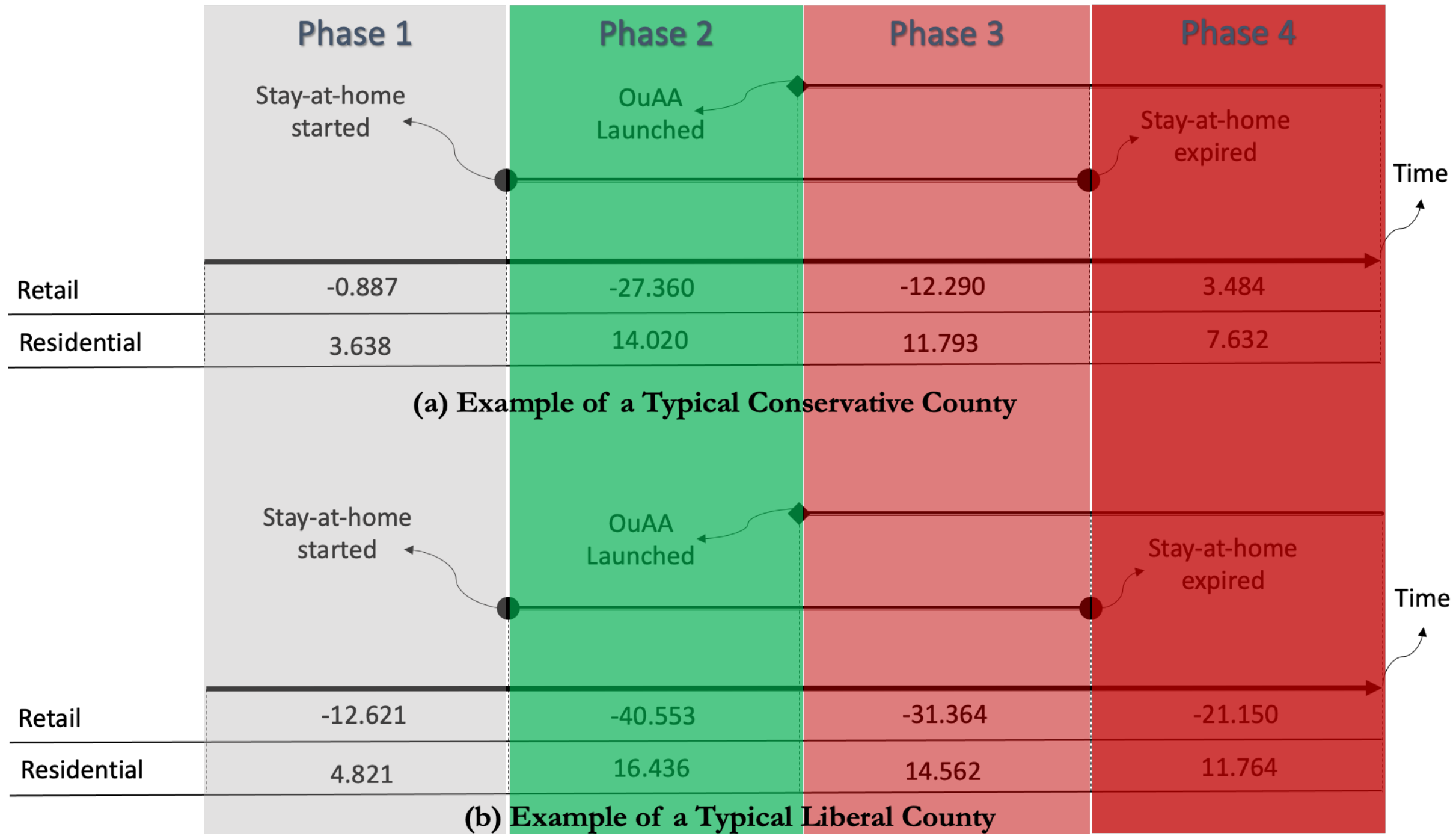

**Figure 6. Predicted Retail and Residential Scores for A Typical Conservative County (a) and A Typical Liberal County (b)**

To better understand the distribution of counties based on whether they enforced a stay-at-home order and whether they did that before the launch of OuAA campaign, we created a Sankey diagram (Figure 7). According to the Sankey diagram, there are 754 liberal counties and 809 conservative counties in our data set. 642 liberal counties and 591 conservative counties enforced stay-at-home orders during our study period. However, 112 liberal counties and 218 conservative counties did not enforce stay-at-home order during our study period. From all the counties that enforced stay-at-home orders (642+591 = 1,233 counties), the majority of them (1,200 counties) enforced stay-at-home orders before the launch of OuAA campaign, and only 33 counties enforced their orders after the launch of OuAA campaign. These numbers indicate that from 1,563 counties in our data set, the majority of them enforced the stay-at-home orders before the launch of OuAA campaign (1,200 counties). That is, the two examples and the four phases we studied in Figure 6 were representative in that they resembled what happened in the majority of the counties during the period of our study.

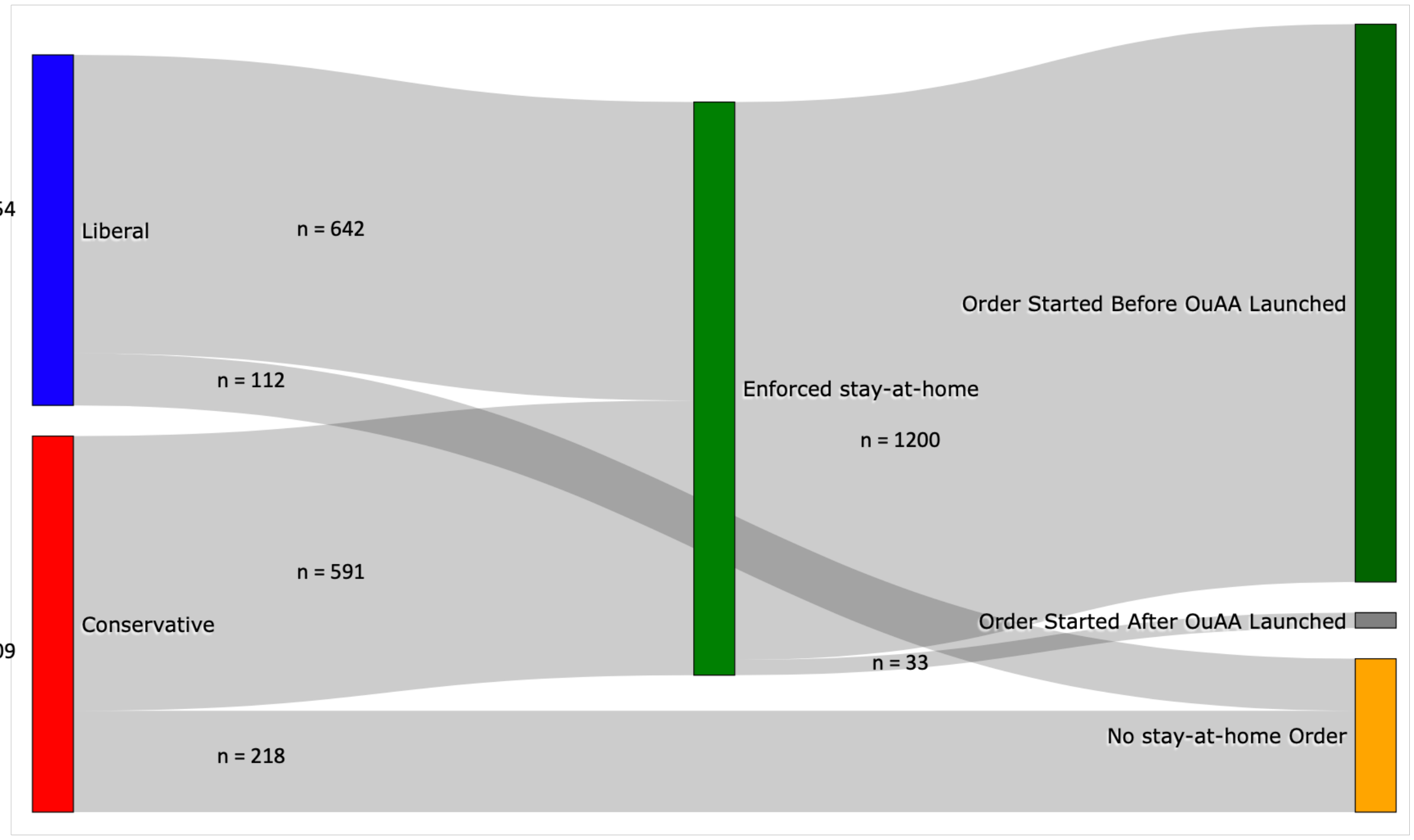

**Figure 7. Sankey Diagram of U.S. Counties Based on Enforcement of Stay-at-home Orders and Launch of OuAA Campaign**

*Robustness Checks:*
Given that the decision to enforce stay-at-home orders could be endogenous, we used propensity score matching to find the best match for each county that had no stay-at-home enforcement with a similar county that enforced state-at-home order in place. We used county level data about residents' education, population, deaths, births, domestic and international migration, percent below federal poverty line, unemployment rate, and median household income to find similar counties (matches for counties that did not have stay-at-home orders). This resulted in 660 counties (330 counties without any stay-at-home order and 330 similar counties that had stay-at-home orders). We repeated our main DiD analysis using this matched sample instead of the entire data. The results of this analysis are reported in Table 4. According to the results reported in this table, our findings are robust.

**Table 4. Results Based on Matched Samples**

| | Retail | Residential |
|---|---|---|
| *Variables* | **Model 7** | **Model 6** |
| Stay | -11.424***<br>[2.058] | 4.684***<br>[0.559] |
| stay × conservative | 13.580***<br>[2.152] | -4.845***<br>[0.585] |
| post_reopen × conservative | 13.908***<br>[1.496] | -2.116***<br>[0.407] |
| stay × post_reopen | 2.190<br>[2.884] | -2.337**<br>[0.784] |
| stay × post_reopen × conservative | -5.925*<br>[2.662] | 1.744*<br>[0.723] |
| cases | -0.068***<br>[0.008] | 0.020***<br>[0.002] |
| deaths | -0.541*<br>[0.215] | 0.255***<br>[0.058] |
| unemployment | -0.294***<br>[0.054] | 0.158***<br>[0.015] |
| other county restrictions | ✓ | ✓ |
| time fixed effects | ✓ | ✓ |
| county fixed effects | ✓ | ✓ |
| time fixed effects × state fixed effects | ✓ | ✓ |
| Observations | 6,292 | 6,292 |
| $R^2$ | 0.936 | 0.955 |
| F-statistic | 138.989*** | 198.916*** |

## Conclusion:

The immense impact of COVID-19 pandemic on the lives of billions of people has forced authorities to devise response strategies to contain the damage. We created a panel data set of 1,563 U.S. counties by collecting data about weekly community mobility scores, weekly COVID-19 new cases and deaths, ideological orientation of counties, and state's COVID-19 response data (stay-at-home and shelter-in-place restrictions timelines) for the period between the first week of March 2020 and the first week of June 2020.

The enforcements of stay-at-home orders by counties over time created a natural experiment setting that allows the comparison of difference in community mobility before and after enforcing the stay-at-home orders across the counties. We used the Difference-in-Difference (DiD) study design to understand the impact of stay-at-home orders on community mobility. We further address the endogeneity of stay-at-home order decisions using a matched sample of counties (a match between counties that enforced an order and counties that never did).

Our results indicate that stay-at-home orders were effective to some extent in decreasing commute to retail stores and increasing time spent at home. We also find that conservative counties were more likely to ignore the stay-at-home orders. This finding is aligned with similar studies about partisan behavior in obeying coronavirus restrictions (Kushner Gadarian et al. 2020). We further find that the "Opening up America Again" (OuAA) campaign launched by The White House increased retail & recreation activities and decreased time spent at home. We also find that in conservative counties that enforced stay-at-home, OuAA campaign was less effective when compared to conservative counties without stay-at-home orders. These results suggest promising news for local authorities. That is, even when the federal government is reopening the country, the local authorities that enforced stay-at-home restrictions were to some extent effective in decreasing the commute to retail stores and recreational facilities such as gyms and increasing time spent at home. Our findings extend the findings of previous research (Alashoor et al. 2020; Grossman et al. 2020).